 \documentclass[a4paper,aps,prd,showpacs,nofootinbib,preprintnumbers,amsmath,amssymb,mcite,superscriptaddress,12pt]{revtex4-1}

\usepackage[table]{xcolor}
\usepackage{graphicx}
\usepackage{dcolumn}
\usepackage{bm}
\usepackage{esvect}
\usepackage{accents}

\usepackage{hyperref}
\hypersetup{colorlinks=true,linkcolor=magenta,anchorcolor=green,citecolor=cyan,filecolor=black,menucolor=black,urlcolor=brown}

\newcommand{\be}{\begin{equation}}
\newcommand{\ee}{\end{equation}}
\newcommand{\ba}{\begin{eqnarray}}
\newcommand{\ea}{\end{eqnarray}}

\newcommand{\Mpl}{M_{\textrm{Pl}}}

\begin{document}

\preprint{IPHT-t18/004}

\title{New Bounds on  Dark Energy Induced Fifth Forces}

\author{Philippe Brax}
\affiliation{Institut de Physique Th\'eorique, Universit\'e  Paris-Saclay, CEA, CNRS, F-91191 Gif-sur-Yvette Cedex, France}
\author{Patrick Valageas}
\affiliation{Institut de Physique Th\'eorique, Universit\'e  Paris-Saclay, CEA, CNRS, F-91191 Gif-sur-Yvette Cedex, France}
\author{Pierre Vanhove}
\affiliation{Institut de Physique Th\'eorique, Universit\'e  Paris-Saclay, CEA, CNRS, F-91191 Gif-sur-Yvette Cedex, France}
\affiliation{National Research University Higher School of Economics, Russian Federation}

\begin{abstract}

  We consider the gravitational Wilsonian effective action at low
  energy when all the particles of the standard model have
  decoupled. When the ${\cal R}^2$ terms dominate, the theory is equivalent
  to a scalar-tensor theory with the universal coupling $\beta=1/\sqrt 6$
  to matter for which we present  strong lower and upper bounds on the
  scalaron mass $m$ obtained by using results from the E\"ot-Wash
  experiment  on the modification of the  inverse-square law, the  observations of the hot gas
  of  galaxy clusters  and the Planck
  satellite data on the neutrino masses. In terms of the range of the scalar
  interaction mediated over a distance of order $m^{-1}$, this leads
  to the small interval $4\,\mu m \lesssim m^{-1} \lesssim 68\, \mu m$
   within reach of future experimental tests of
  deviations from Newton's gravitational inverse-square law.

\end{abstract}

\date{\today}

\maketitle


\section{Introduction}
\label{sec:introduction}

The recent results by LIGO/VIRGO show that most of the self-acceleration models of dark
energy and modified gravity lead to excluded speeds for the gravitational waves~\cite{TheLIGOScientific:2016src,Abbott:2017vtc}.
This leaves only as a viable option a whole swath of dark energy models where the magnitude
of the vacuum energy has to be tuned. A possibility which has not been explored so far
is that gravity itself, seen as a low energy effect, could actively participate in the mechanism leading to the acceleration without
new degrees of freedom being added in an ad hoc fashion (such as quintessence fields
or additional metrics).
This can be realized using  the higher order corrections to the Einstein-Hilbert action, which contain
higher derivative interactions. Indeed, the fundamental symmetries of General
Relativity allow for the presence of  these higher derivative contributions
to the local and diffeomorphism-invariant Lagrangian.  Even if they are classically set to zero, they are
generated by quantum corrections as counter-terms to ultraviolet
divergences~\cite{tHooft:1974toh,Goroff:1985th}. These corrections can be seen to generate  new intrinsic
degrees of freedom. Generically, these degrees of freedom are ghost-like and can only be
tolerated at low energy when the suppression scale of all the higher order operators lies at
the cut-off scale of the low energy
description~\cite{Burgess:2007pt}.

As the dark energy
scale $\rho_{\rm vac}=3\Omega_\Lambda M_{\rm Pl}^2 H_0^2 $ is of
the milli-electron-Volt mass order, $\rho_{\rm vac}^{1/4} \simeq 2$~meV, we can also integrate out the electron and  all heavier particles of
the standard model and concentrate on the very low energy degrees of freedom when considering the dynamics of the Universe at late time. This sets the cut-off scale of the low-energy effective field theory
to  the mass of the electron as a typical order of magnitude.
The remaining low-energy degrees of freedom are  gravity itself and the neutrinos.
Amongst the higher derivative interactions in the gravitational sector, the Ricci scalar squared
${\cal R}^2$ invariant plays a special role as  it does not give rise  to ghosts and is the most relevant interaction at low energy  amongst the higher order gravitational
interaction terms.
This motivates hierarchical scenarios where the scalaron associated
with ${\cal R}^2$ has a low energy mass  while the ghost-like contributions
are rejected  at the cut-off scale, as they should to satisfy theoretical and
observational constraints.
Then, the quantum fluctuations of the scalaron could act as  dark energy and lead to the acceleration of the
expansion of the Universe, without invoking any new physics at low
energy.
We have explicitly presented in~\cite{Brax:2017bcp} such a model,
where the scalaron provides the new light degree of freedom that sets
the dark energy scale and keeps a local quantum field description, while
the ghost-like contributions only arise at the cut-off scale.

In this paper we relax this hypothesis and do not consider
  that dark energy is necessarily driven by the scalaron's vacuum
  energy only. We also allow for the existence of other light and
  decoupled-from-matter scalar degrees of freedom, such as light
  bosonic dark matter, of masses smaller than $ 0.1$ eV, which can
  also contribute to the dark energy. In their absence we predict that
  the scalaron's mass must be close to the averaged neutrino mass
  $\bar m_\nu$ (see section~\ref{sec:neutr-contr}) of
  order $0.1$ eV in order to compensate for the negative contribution
  of the neutrinos to the dark energy and to satisfy the bound on the
  vacuum energy density in galaxy clusters. In this case, { as we
  explain,} the scalaron
  { induces} a deviation from Newton's law with a coupling strength
  $\beta=1/\sqrt{6}$ at a distance of order $4~\mu$m. On the other
  hand if decoupled light scalars of masses less than $0.1$ eV are
  present, they { must} evade gravitational tests and their contributions to
  the dark energy together with the scalaron's compensate for the
  neutrino's. This implies strong bounds~(\ref{range}) on the scalaron
  mass $m$ close to the current sensitivity of the E\"ot-Wash
  experiments~\cite{Hoyle:2004cw,Kapner:2006si} and within reach of
  the new runs~\cite{AdelbergerTalk} which have been recently
  presented.

\vspace{-.5cm}
\section{The low energy effective $\mathcal R^2$ model}
\label{sec:mathcal-r2-model}
\vspace{-.3cm}

Our metric has signature $(- + + + )$.
At low energy, below a cut-off scale $M$,  the leading correction to the Einstein-Hilbert action reads $S + \delta S$, where
\begin{equation}\label{e:SR2}
S=\int d^4x \sqrt{-g}\Big[ \frac{\Mpl^2}{2} {\cal R} - \rho_{\Lambda}(\mu)
+  c_0 (\mu) {\cal R}^2+ c_2 (\mu)
                                      (R_{\mu\nu}R^{\mu\nu}-\frac{1}{3}
                                      {\cal R}^2) \Big] ,
\end{equation}
and $\delta S$ contains all the higher order terms in the curvature
invariants,
\begin{equation}
\delta S=\int d^4x \sqrt{-g}\frac{\Mpl^2 }{2}\sum_{n\ge 3}
\alpha_n(\mu) M^2\left(R\over M^2\right)^n\, .
\end{equation}
Here $\alpha_n (\mu)$ are dimensionless coefficient of order
${\cal O}(1)$ and $R$ stands for the various Riemann tensor components
$R_{\mu\nu\rho\sigma}$.  The scale $M$ plays the role of the cut-off
scale of the effective gravitational field theory, which is valid for
$R\ll M^2\ll \Mpl^2$.   We are interested in a low-energy
  effective action in the energy range of the dark energy scale, below
  the electron mass, $\mu\ll m_e$.  In this regime the typical
  curvature is tiny on cosmological
  scales~\cite{Brax:2016vpd}.  As an effective field theory with higher order
  derivatives, there are new degrees of freedom per power of the
  Riemann tensor.  They are generically ghost-like with a mass of
  order the Ultra-Violet cut-off scale ${\cal O}(M)$. Therefore, they do not play a role
  at low energy below $M$.

 It has been shown in~\cite{Fradkin:1981iu,Salvio:2014soa} that the coefficient $c_2(\mu)$ is always asymptotically free, since
$dc_2(\mu)/d\log \mu^2>0$, whereas $c_0(\mu)$ is asymptotically safe,
$dc_0(\mu)/d\log\mu^2<0$. For the non tachyonic case $c_0(\mu)>0$, which
is the case of interest in this paper.
Therefore, at low-energy $c_2(\mu)$ tends to zero whereas $c_0(\mu)$
grows,  leading to  the hierarchy $c_0(\mu)\gg c_2(\mu)$ at very
low energy.  Hence the quadratic Ricci scalar term is  enhanced as compared with other quadratic and higher order  contributions.

We will then work with the low-energy effective action
\begin{equation}\label{e:SR2bis}
S_\mu=\int d^4x \sqrt{-g} \left[ \frac{\Mpl^2}{2} {\cal R} - \rho_{\Lambda}(\mu)+  c_0(\mu) {\cal R}^2 \right] + S_{\rm matter} \, ,
\end{equation}
obtained
after integrating out all the massive particles of the standard model
of masses above the electron mass, i.e. the matter action only involves the light matter fields of masses less than the electron mass.

\section{The minimal scalaron  model}

The ${\cal R}^2$ theories  are equivalent to  scalar field models
 as reviewed in~\cite{Sotiriou:2008rp}. As such, they also correspond to the
 scalar-tensor theories
\begin{equation}\label{e:SR2ter}
S_{\mu,\phi}=\int d^4x \sqrt{-g}\left ( \frac{\Mpl^2}{2} {\cal R} -
  \rho_{\Lambda}(\mu) - \frac{(\partial \phi)^2}{2} -\frac{m^2}{2}
  \phi^2\right ) + S_{\rm matter} (\psi_i, e^{2\beta\phi/\Mpl} g_{\mu\nu}) .
\end{equation}
The massive scalar couples  to the matter
stress-energy tensor  with the universal strength $\beta=1/\sqrt{6}$ and the mass of the scalaron is
\begin{equation}
  m(\mu)^2= {\beta^2\Mpl^2\over 2c_0(\mu)}  \;\;\; \mbox{with} \;\;\;
  \beta= \frac{1}{\sqrt{6}} ,
\end{equation}
and
$S_{\rm matter}$ is the matter action depending on the matter fields
$\psi_i$.
The scalar potential has been expanded to lowest order in
$\phi/\Mpl$ as we are interested in the regime where
$c_0{\cal R} \ll \Mpl^2$, and the standard Einstein-Hilbert term dominates
so as to ensure convergence to General Relativity in the very low
curvature regime.  The scalaron self-interactions are
  negligible, being suppressed by powers of $m$ which is very small in
  Planck mass units.

The mass of the scalaron is affected by renormalisation effects. In the Jordan frame, where the particles of the standard models are coupled
to $e^{2\beta\phi/\Mpl} g_{\mu\nu}$,   the quantum fluctuations due to massive particles and the phase transitions in the matter sector are all scalar-independent. It is only when changing to the Einstein frame that
the potential term of the $\mathcal R^2$ model is corrected by a term $e^{4\beta\phi/\Mpl}\rho_\Lambda (\mu)$ coming from the coupling of the energy density to matter. Expanding $e^{4\beta\phi/\Mpl}$ in powers of $\phi/M_{\rm Pl}$ leads to the $\mu$ dependent mass
\begin{equation}
m^2(\mu)= m^2 + 16\beta^2 \frac{\rho_\Lambda(\mu)}{M^2_{\rm Pl}} .
\end{equation}
At low energy below the electron mass, this correction is negligible as
we shall confirm below. This implies that the coefficient $c_0$ in the
low-energy $\mathcal R^2$ action is also independent of renormalisation
effects associated with the massive fields of the standard model.

To sum-up, the low-energy degrees of freedom below the electron mass
are the neutrinos and the scalaron  of mass $m \ll m_e$, associated with the Ricci scalar
${\cal R}^2$ term.  Notice that the
new cut-off of this effective action is $m_e$.

\section{Cosmological constant and vacuum energy}

We consider this effective field theory to be valid at a scale $\mu$ much lower than the
electron mass, and to describe the late-time acceleration of the expansion of the
Universe.  At this energy scale, the only fields which
have not been integrated out yet, when  less massive than $\mu$, are the three massive
neutrinos and the scalaron $\phi$.
The vacuum energy $\rho_\Lambda(\mu)$ corresponds to the
combined effect of all the quantum corrections associated to massive
particles that have been integrated out, the various phase
transitions including the QCD and electro-weak ones  ${\rho^{SM}_\Lambda(\mu)}$, and the bare
cosmological constant $ \rho_\Lambda^0$ seen as finite counter-term after renormalisation,
\begin{equation}\label{e:defrhoL}
\rho_\Lambda(\mu)= \rho^{SM}_\Lambda(\mu)+\rho_\Lambda^0\,.
\end{equation}
We shall work in the minimal decoupling subtraction scheme $\overline{DS}$~\cite[\S4.1.4]{Burgess:2007pt}, where the parameters of the Wilsonian action
can be derived by integrating the renormalisation group equations
obtained by taking into account in the $\beta$ functions only the
particles that have not been integrated out yet. We refer to~\cite{Burgess:2007pt} for a lucid presentation of
the Wilsonian  effective actions   and~\cite{Sola:2013gha} for a review of various
renormalisation group
approaches to the cosmological constant question. Thus, at energies
$\mu <m_e$, {at one-loop order the} vacuum energy
receives quantum corrections due to the scalar field $\phi$ and the
neutrinos only. This leads to the  renormalisation group equation
\begin{equation}\label{e:drho}
  {d\rho_\Lambda(\mu)\over d\log\mu^2}= -{m^4\over
    64\pi^2} \theta(\mu>m)
    +2\sum_{f=1}^3{m_f^4\over 64\pi^2} \theta(\mu>m_f)   \,,
\end{equation}
which includes the contribution of the three neutrinos, see the appendix~A of \cite{Brax:2016jjt} for the one-loop evaluation.
The Heaviside factors ensure that the scalaron and the neutrinos
no longer contribute to the running at energies below their mass threshold,
as they decouple \cite{Burgess:2007pt}.
There are no contributions from photons or gluons in this
very low energy regime. A similar equation has already been considered in~\cite{Shapiro:1999zt}
without the scalaron contribution.

At low energy, $\max(m,m_f) < \mu <m_e$, before the scalaron and the
neutrinos decouple, the vacuum energy $\rho_\Lambda (\mu)$ is given by
\begin{equation}
\label{e:rho}
\rho_\Lambda(\mu)= \rho_{\rm \Lambda}(m_e) + \left(\frac{m^4}{64\pi^2}  -2\sum_{f=1}^3
\frac{m_f^4}{64\pi^2} \right)\ln \frac{m_e^2}{\mu^2}
\end{equation}
It matches with the vacuum energy $\rho_\Lambda (m_e)$
at  the energy scale $\mu=m_e$, coming from the evolution of $\rho_\Lambda (\mu)$
at energies $\mu >m_e$ as required by
  the decoupling~\cite{Georgi:1994qn} at the scale  the electron mass.
On the other hand,  when considering dark energy,
as the Hubble scale today is $H_0 \sim 10^{-42} {\rm GeV}$,
we are interested in the very  low energy Wilsonian action  for $\mu$ much lower than the  neutrino masses and the scalaron mass. In this regime,
the neutrinos and the scalaron have
decoupled and
the vacuum energy becomes a constant corresponding to the 1PI vacuum energy which appears in the classical equations of motion of the theory.
This is  the dark energy density $\rho_{\rm vac}$ measured by
cosmological probes,
\begin{equation}
\rho_{\rm vac} \simeq 2.7 \times
10^{-11}~\textrm{eV}^4.
\label{pres}
\end{equation}
Therefore, integrating~(\ref{e:drho}) from $m_e$ down to $\mu < \min(m,m_f)$,
taking into account the jump of the vacuum energy $\beta$ function at the masses of the
scalaron and the neutrinos, we get
\begin{equation}
\rho_{\rm vac}= \rho_{\rm \Lambda}(m_e) + \frac{m^4}{64\pi^2} \ln
\frac{m_e^2}{m^2} - 2\sum_{f=1}^3
\frac{m_f^4}{64\pi^2} \ln \frac{m_e^2}{m_f^2} .
\label{e:rho-vac-me}
\end{equation}

The energy density $\rho_{\rm vac}$ is much lower than
the order of magnitude of particle physics scales, e.g.  early Universe phase transitions and quantum fluctuations of very massive fields.
These contributions to the vacuum energy density  have all been subsumed in
$\rho_\Lambda (m_e)$, which contains all the physical effects
at energies higher than $m_e$ and contributing to the renormalised energy density, where the bare cosmological constant has been used as
a counter-term in the renormalisation process.

\section{Neutrino contributions}
\label{sec:neutr-contr}

We note that the neutrino contributions to $\rho_{\rm vac}$ are  strongly constrained by
cosmological and astrophysical measurements.
The Planck results~\cite{Aghanim:2018eyx,Giusarma:2016phn,Vagnozzi:2017ovm,Giusarma:2018jei} for the cosmic microwave background provide
the upper bound $m_1+m_2+m_3 < 0.12~$eV for the sum of the neutrino
masses.\footnote{We would like to thank Sunny Vagnozzi for
  communications 
  about the most recent  bound on the neutrino masses from Planck data.}
The oscillations of the solar neutrinos yield the squared mass difference
$m_2^2  - m_1^2=7.5\,10^{-5}~$eV$^2$.
For the case of normal ordering of neutrino masses~\cite{Esteban:2016qun}, $m_3^2-m_1^2=2.524\,10^{-3}~\textrm{eV}^2$,
whereas for the case of inverse ordering, $m_3^2-m_2^2=2.514\,10^{-3}\,\textrm{eV}^2$.
For both orderings the neutrino contribution is bounded,
\begin{equation}\label{e:NeutrinoBounds}
10^4\,\rho_{\rm vac} \leq \sum_{f=1}^3  {m_f^4\over
  64\pi^2}\,\log\left(m_e^2\over m_f^2\right)\leq 2\times
10^4\rho_{\rm vac}\,.
\end{equation}
This is of course a remnant of the usual cosmological constant problem, i.e. the
overestimate of the vacuum energy density by particle physics expectations.
It implies that either the vacuum energy $\rho_\Lambda (m_e)$,
the scalaron contribution in $m^4$, or their sum, must compensate the
neutrino contribution
\begin{equation}-{\bar m_\nu^4\over 32\pi^2} \ln(m_e^2/\bar m_\nu^2)=
  -\sum_{f=1}^{3}{m_f^4\over32\pi^2}\ln(m_e^2/m_f^2)
\end{equation}
in~(\ref{e:rho-vac-me}).

\vspace{-.5cm}
\section{Bounds on the scalaron mass }
\label{sec:constraints}
\vspace{-.3cm}

\medskip
\noindent{\bf Lower bound from E\"ot-Wash experiments.}
The scalar curvature square term $\mathcal R^2$ induces a
  modification of the large-distance gravitational potential from
  objects of mass $M$~\cite{Stelle:1977ry},
  \begin{equation}\label{e:Vpot}
    V(r) = -{GM\over r}\, \left( 1+{1\over3}e^{-mr}\right)    \,.
  \end{equation}
 The absence of evidence for  short range forces  in the
E\"ot-Wash experiment~\cite{Hoyle:2004cw,Kapner:2006si} provides a strong
upper bound on the range of such fifth forces,
\begin{equation}
\label{Eot-Wash-range}
m^{-1} \lesssim 68\, \mu m\,.
\end{equation}
This also reads
\begin{equation}\label{e:lower-bound}
m\gtrsim 2.8\,10^{-3}~{\rm eV} \simeq 1.22 \rho_{\rm vac}^{1/4},
 \end{equation}
 which happens to be
of the same order as $\rho_{\rm vac}^{1/4}$.
As pointed out in \cite{Brax:2017bcp}, it is thus possible that the scalaron
would be responsible for the vacuum energy density observed today in~(\ref{e:rho-vac-me}), in which case its mass would be close to current
E\"ot-Wash experimental bounds. However, in this paper we do not assume
that this is the case and relax the link between the scalaron's quantum fluctuations and the vacuum energy.

\medskip
\noindent{\bf Constraints from galaxy clusters.}
The X-ray emitting gas of a galaxy cluster has
a typical temperature of $T_{\rm X} \sim 1$ keV, in regions of total
baryonic and dark matter density of about 500 times the mean density
of the Universe, i.e. $200\rho_{\rm vac}$. These systems typically appeared at a redshift $z \gtrsim 0.1$
and already have a lifetime of the order of the age of the Universe.
In such clusters the scalaron and the neutrinos,  coming either from the early Universe with an energy of order $10^{-4}$ eV or from astrophysical processes such as the burning of stars
with an energy around $100$ keV, have a very small cross section with matter $\sigma \simeq \beta^2/M_{\rm Pl}^2$ and $\sigma \simeq m_e^2/M^4_Z$, where $M_Z\simeq 10^2$ GeV is the mass of the $Z$ boson,  respectively.
This implies that both the neutrinos and the scalaron decouple from the physics inside the clusters,
 which can then be described by  non-relativistic matter particles (such as electrons and protons)
and General Relativity augmented with a vacuum energy.
We make the strong assumption that the latter only  takes into account all the physics for energy scales greater than $T_X$, i.e. $\rho_\Lambda (m_e)$, which excludes the quantum fluctuations of the
scalarons and the neutrinos which have decoupled from the plasma.
This implies strong constraints on the scalaron mass.
Indeed, if the vacuum energy $\vert\rho_\Lambda (m_e)\vert$ were greater than the local matter density
within the virial radius, where the gas has been shocked to $T_{\rm X} \sim 1$ keV,
it would significantly affect the dynamics within the cluster.
In a spherical approximation, the cluster would behave as a separate universe
\cite{Weinberg:1987dv}, with its own vacuum energy $\rho_\Lambda (m_e)$.
The formation of the cluster, from the turn-around time until virialization, would proceed in the same
manner, but the later stages when the gas reaches high temperatures
would be such that the hydrostatic equilibrium would be displaced or beyond reach.
To ensure small dynamical effects, we have the conservative bound
 \begin{equation}
  | \rho_\Lambda (m_e) |   \lesssim 200 \,  \rho_{\rm vac} ,
   \label{lower}
 \end{equation}
 as the hot gas is  typically measured in X-ray clusters at density
 contrasts of 500 compared to the cosmological background.
This reasoning makes use of the presence at low redshift of hot high-density
structures within the cooler and lower-density cosmological background.
Combining (\ref{lower}) with the neutrino bounds~(\ref{e:NeutrinoBounds}),
we obtain from~(\ref{e:rho-vac-me})
the numerical estimate $m \simeq \bar m_\nu\simeq 0.05$~eV, hence we get the noticeably short range $m^{-1} \simeq 4\, \mu m$,
 which is compatible with the E\"ot-Wash bound (\ref{Eot-Wash-range}).

\medskip
\noindent{\bf The extended scalaron model}
 So far we have assumed that only the scalaron and the neutrinos have a mass smaller than the electron mass. Other
particles such as light bosonic dark matter candidates~\cite{Agrawal:2017eqm}  could also be present. In this extended scenario, the bounds (\ref{lower}) could be satisfied as long as the scalaron's and the other scalar fields' contributions to the vacuum energy almost compensate the one of the neutrinos. In this case the mass of the scalaron is still bounded from above by $\bar m_\nu$ whilst being bounded from below by the E\"ot-Wash bound $1.22 \rho_{\rm vac}^{1/4}$ leading to a range
\begin{equation}
4\,\mu m \lesssim m^{-1} \lesssim 68\, \mu m\,.
\label{range}
\end{equation}
for the scalaron-mediated interaction.

\vspace{-.5cm}
\section{Laboratory experiments }
\label{sec:lab}
\vspace{-.3cm}

Different types of experiments can in principle test new interactions in the range
of a few micrometers.
Preliminary results of new runs of
this experiment~\cite{AdelbergerTalk} indicate a possible
new upper bound of $40~\mu m$, therefore reducing the range of allowed mass
almost  by half.
Another experiment which could test the presence of a scalaron is the measurement of the
energy levels of the neutron over a horizontal mirror at $z=0$  in the terrestrial gravitational field~\cite{Nesvizhevsky:2002ef}. The presence of the scalaron
would shift the n-th energy level $\vert n\rangle$ by an amount
\begin{equation}
\delta E_n= -\alpha_n \frac{\beta^2 m_N \rho}{M^2_{\rm Pl} m^2} e^{-m z_0}
\end{equation}
where $m_N$ is the neutron mass, $\rho$ the density of the mirror and the ${\cal O}(1)$ number $\alpha_n$ is such that
$\langle n\vert e^{-mz}\vert n\rangle= \alpha_n e^{-mz_0}$ where $z_0=(\hbar ^2/2m_N^2 g)^{1/3}\simeq 6 \ {\rm \mu m}$. Detecting a scalaron of mass $m\lesssim z_0^{-1}$ for $\rho\simeq 10\ {\rm g/cm^3}$ would require
to have a sensitivity on the energy levels of order $10^{-22}$ eV. This is much below the present sensitivity of order $10^{-14}$ eV~\cite{Jenke:2014yel}, and even below the best sensitivity achievable by such an experiment which
is given by the inverse of the neutron life-time thanks to the uncertainty relation $\Delta E \simeq 10^{-19}$ eV. Hence this type of experiment will never be sensitive to the scalaron.
 Casimir force experiments could lead to strong constraints on Yukawa
  exponential corrections $\alpha\,e^{-r/\lambda}/r$ to the
  Newton potential~\cite{Fischbach:1999bc}. The scale~\eqref{range} would be eventually tested if Yukawa interactions were probed in the 10 micrometer ballpark. This would
correspond to future Casimir experiments such as CANNEX~\cite{Almasi:2015zpa} but their expected sensitivity at the $0.1\ {\rm pN/cm^2}$ would not be low enough
to compete with direct searches for gravitational interactions. Indeed the scalar pressure between two plates separated by a distance $d$ is given by
\begin{equation}
\frac{F}{A}=  \frac{\beta^2 \rho^2}{2M^2_{\rm Pl} m^2} e^{-m d}.
\end{equation}
For a scalaron of mass given by the E\"ot-Wash bound and a distance of 10 microns, this would require a sensitivity of around $10^{-2} {\rm pN/cm^2}$  which is one order of magnitude
below the CANNEX expected sensitivity. For larger masses corresponding to short ranges for the scalaron interaction, the sensitivity would have to be even better. Finally
 one extremely promising possibility which would overcome
some of the shortcomings of ground-based experiments would be to have a torsion pendulum experiment of the E\"ot-Wash type aboard a satellite.
Such a project has already been considered~\cite{island1} with a target of force ranges around 10 micrometers which would be sensitive to coupling of order one or below
such as $\beta = 1/\sqrt 6$~\cite{island2}. Of course such a future experiment would have the power to vindicate or exclude the scalaron that we have considered in this work as the torque between two parallel and rotating plates of common surface area $A(\theta)$ depending on the rotation angle $\theta$  is given by
\begin{equation}\label{e:Tmeasure}
T= \frac{dA(\theta)}{d\theta} \,\frac{\beta^2 \rho^2}{2 M_{\rm Pl} m^3} e^{-md} \, ,
\end{equation}
where $\rho$ is their common density and $d$ their separation. Measurements of the Yukawa decrease and the amplitude would give access to both $m$ and $\beta$.

\vspace{-.5cm}
\section{Discussion}
\vspace{-.4cm}

Upon the hypothesis that inside clusters of galaxies the relevant vacuum energy density
is given by $\rho_\Lambda(m_e)$, we have bounded the mass of the scalaron
giving  rise to a modification of the Newtonian
potential~(\ref{e:Vpot}) with a range
within reach of the new runs~\cite{AdelbergerTalk} of the E\"ot-Wash experiment.
The measurement of the value of coupling to matter $\beta$  compatible with $1/\sqrt6$
would point towards  a gravitational theory $f(\mathcal R)$ with
${\cal R}^2$ being the leading contribution~\cite{Brax:2017bcp}.  An altogether different value would indicate that the detected scalar
is not the one generated by gravitational corrections to General Relativity and would come from some new and unknown physics at low energy.
If no signal in laboratory experiments searching for fifth forces were
found in this small mass range, this would also signify that the scalaron has a
much larger mass  with a much smaller coefficient $c_0$ closer to the one of
the other curvature squared terms. This would {eventually} imply that the hierarchy
$c_0\gg c_2$ is not realised and that
new light
degrees of freedom as suggested in~\cite{Banks:1988je,Beane:1997it} must be present at low energy. Such degrees of
freedom would have to be very weakly coupled to have escaped
direct detection. In this case these light degrees of freedom  would have their masses
bounded from above by the averaged  neutrino mass $\bar m_\nu\simeq 0.05 $ eV.

\section{Acknowledgements}

We would like to thank Andrei Barvinsky, Cliff Burgess, Gregory Korchemsky,
St\'ephane Lavignac, Subodh Patil, Stefan Theisen, Mikhael
Vasiliev, Sunny Vagnozzi and Boris Voronov for useful discussions.  The research of P. Vanhove has
received funding the ANR grant ``Amplitudes'' ANR-17- CE31-0001-01,
and is partially supported by Laboratory of Mirror Symmetry NRU HSE,
RF Government grant, ag. N$^\circ$ 14.641.31.0001.  This work is
supported in part by the EU Horizon 2020 research and innovation
programme under the Marie-Sklodowska grant No. 690575. This article is
based upon work related to the COST Action CA15117 (CANTATA) supported
by COST (European Cooperation in Science and Technology).


\begin{thebibliography}{99}
  
\bibitem{TheLIGOScientific:2016src}
B.~P.~Abbott {\it et al.} [LIGO Scientific and Virgo Collaborations],
``Tests of General Relativity with GW150914,''
Phys.\ Rev.\ Lett.\ {\bf 116} (2016) no.22, 221101
[arXiv:1602.03841].

\bibitem{Abbott:2017vtc}
B.~P.~Abbott {\it et al.} [LIGO Scientific and VIRGO Collaborations],
``GW170104: Observation of a 50-Solar-Mass Binary Black Hole Coalescence at Redshift 0.2,''
Phys.\ Rev.\ Lett.\ {\bf 118} (2017) no.22, 221101
[arXiv:1706.01812].


\bibitem{tHooft:1974toh}
G.~'t Hooft and M.~J.~G.~Veltman,
``One Loop Divergencies in the Theory of Gravitation,''
Ann.\ Inst.\ H.\ Poincare Phys.\ Theor.\ A {\bf 20} (1974) 69.

\bibitem{Goroff:1985th}
M.~H.~Goroff and A.~Sagnotti,
``The Ultraviolet Behavior of Einstein Gravity,''
Nucl.\ Phys.\ B {\bf 266} (1986) 709.

\bibitem{Burgess:2007pt}
C.~P.~Burgess,
``Introduction to Effective Field Theory,''
Ann.\ Rev.\ Nucl.\ Part.\ Sci.\ {\bf 57} (2007) 329
[hep-th/0701053].

\bibitem{Brax:2017bcp}
P.~Brax, P.~Valageas and P.~Vanhove,
``$R^2$ Dark Energy in the Laboratory,''
Phys.\ Rev.\ D {\bf 97} (2018) no.10, 103508
[arXiv:1711.03356 [astro-ph.CO]].


\bibitem{Hoyle:2004cw}
C.~D.~Hoyle, D.~J.~Kapner, B.~R.~Heckel, E.~G.~Adelberger, J.~H.~Gundlach, U.~Schmidt and H.~E.~Swanson,
``Sub-Millimeter Tests of the Gravitational Inverse-Square Law,''
Phys.\ Rev.\ D {\bf 70} (2004) 042004
[hep-ph/0405262].

\bibitem{Kapner:2006si}
D.~J.~Kapner, T.~S.~Cook, E.~G.~Adelberger, J.~H.~Gundlach, B.~R.~Heckel, C.~D.~Hoyle and H.~E.~Swanson,
``Tests of the Gravitational Inverse-Square Law Below the Dark-Energy Length Scale,''
Phys.\ Rev.\ Lett.\ {\bf 98} (2007) 021101
[hep-ph/0611184].


\bibitem{AdelbergerTalk} E. Adelberger, ``Probing quantum-gravity
  notions via classical tests of  the Equivalence Principle and the
  Inverse-square Law'', Talk at The 13th Central European Seminar on Particle Physics and Quantum Field Theory, November 30 - December 1, 2017, University of Vienna.

\bibitem{Brax:2016vpd}
P.~Brax, L.~A.~Rizzo and P.~Valageas,
``Ultralocal Models of Modified Gravity without Kinetic Term,''
Phys.\ Rev.\ D {\bf 94} (2016) no.4, 044027
[arXiv:1605.02938 [astro-ph.CO]].


\bibitem{Fradkin:1981iu}
E.~S.~Fradkin and A.~A.~Tseytlin,
``Renormalizable Asymptotically Free Quantum Theory of Gravity,''
Nucl.\ Phys.\ B {\bf 201} (1982) 469.


\bibitem{Salvio:2014soa}
A.~Salvio and A.~Strumia,
``Agravity,''
JHEP {\bf 1406} (2014) 080
[arXiv:1403.4226 [hep-ph]].

\bibitem{Sotiriou:2008rp}
T.~P.~Sotiriou and V.~Faraoni,
``F(R) Theories of Gravity,''
Rev.\ Mod.\ Phys.\ {\bf 82} (2010) 451
[arXiv:0805.1726].



\bibitem{Sola:2013gha}
J.~Sola,
``Cosmological Constant and Vacuum Energy: Old and New Ideas,''
J.\ Phys.\ Conf.\ Ser.\ {\bf 453} (2013) 012015
[arXiv:1306.1527 [gr-qc]].

\bibitem{Brax:2016jjt}
  P.~Brax and P.~Valageas,
  ``Quantum Field Theory of K-Mouflage,''
  Phys.\ Rev.\ D {\bf 94} (2016) no.4,  043529
  [arXiv:1607.01129 [astro-ph.CO]].

\bibitem{Shapiro:1999zt}
I.~L.~Shapiro and J.~Sola,
``On the Scaling Behavior of the Cosmological Constant and the Possible Existence of New Forces and New Light Degrees of Freedom,''
Phys.\ Lett.\ B {\bf 475} (2000) 236
[hep-ph/9910462].

\bibitem{Georgi:1994qn}
  H.~Georgi,
  ``Effective field theory,''
  Ann.\ Rev.\ Nucl.\ Part.\ Sci.\  {\bf 43} (1993) 209.

  \bibitem{Aghanim:2018eyx}
  N.~Aghanim {\it et al.} [Planck Collaboration],
  ``Planck 2018 results. VI. Cosmological parameters,''
  arXiv:1807.06209 [astro-ph.CO].

  \bibitem{Giusarma:2016phn}
  E.~Giusarma, M.~Gerbino, O.~Mena, S.~Vagnozzi, S.~Ho and K.~Freese,
  ``Improvement of cosmological neutrino mass bounds,''
  Phys.\ Rev.\ D {\bf 94} (2016) no.8,  083522
  [arXiv:1605.04320 [astro-ph.CO]].

  \bibitem{Vagnozzi:2017ovm}
  S.~Vagnozzi, E.~Giusarma, O.~Mena, K.~Freese, M.~Gerbino, S.~Ho and M.~Lattanzi,
  ``Unveiling $\nu$ secrets with cosmological data: neutrino masses and mass hierarchy,''
  Phys.\ Rev.\ D {\bf 96} (2017) no.12,  123503
  [arXiv:1701.08172 [astro-ph.CO]].

\bibitem{Giusarma:2018jei}
  E.~Giusarma, S.~Vagnozzi, S.~Ho, S.~Ferraro, K.~Freese, R.~Kamen-Rubio and K.~B.~Luk,
  ``Scale-dependent galaxy bias, CMB lensing-galaxy cross-correlation, and neutrino masses,''
  Phys.\ Rev.\ D {\bf 98} (2018) no.12,  123526
  [arXiv:1802.08694 [astro-ph.CO]].


\bibitem{Esteban:2016qun}
  I.~Esteban, M.~C.~Gonzalez-Garcia, M.~Maltoni, I.~Martinez-Soler and T.~Schwetz,
  ``Updated fit to three neutrino mixing: exploring the accelerator-reactor complementarity,''
  JHEP {\bf 1701} (2017) 087
  [arXiv:1611.01514].

\bibitem{Stelle:1977ry}
K.~S.~Stelle,
``Classical Gravity with Higher Derivatives,''
Gen.\ Rel.\ Grav.\ {\bf 9} (1978) 353.


\bibitem{Weinberg:1987dv}
  S.~Weinberg,
  ``Anthropic Bound on the Cosmological Constant",
  Phys. Rev. Lett.  {\bf 59}, 2607 (1987).


\bibitem{Agrawal:2017eqm}
  P.~Agrawal, G.~Marques-Tavares and W.~Xue,
  ``Opening up the QCD axion window,''
  JHEP {\bf 1803} (2018) 049
[arXiv:1708.05008 [hep-ph]].

\bibitem{Nesvizhevsky:2002ef}
  V.~V.~Nesvizhevsky {\it et al.},
  ``Quantum states of neutrons in the Earth's gravitational field,''
  Nature {\bf 415} (2002) 297.

\bibitem{Jenke:2014yel}
  T.~Jenke {\it et al.},
  ``Gravity Resonance Spectroscopy Constrains Dark Energy and Dark Matter Scenarios,''
  Phys.\ Rev.\ Lett.\  {\bf 112} (2014) 151105
[arXiv:1404.4099 [gr-qc]].

\bibitem{Fischbach:1999bc}
  E.~Fischbach and C.~L.~Talmadge,
  ``The search for nonNewtonian gravity,''
  New York, USA: Springer (1999).


\bibitem{Almasi:2015zpa}
  A.~Almasi, P.~Brax, D.~Iannuzzi and R.~I.~P.~Sedmik,
  ``Force sensor for chameleon and Casimir force experiments with parallel-plate configuration,''
  Phys.\ Rev.\ D {\bf 91} (2015) no.10,  102002
  [arXiv:1505.01763].

\bibitem{island1} J. Berg\'e, ``The Inverse Square Law And Newtonian Dynamics space explorer (ISLAND)'', 2017,  Proceedings of the 52nd Rencontres de Moriond.

\bibitem{island2} J. Berg\'e, ``The Inverse Square Law And Newtonian Dynamics space explorer (ISLAND)'', 2017, http://moriond.in2p3.fr/grav/2017/program.php.

\bibitem{Banks:1988je}
T.~Banks,
``Prolegomena to a Theory of Bifurcating Universes: a Nonlocal Solution to the Cosmological Constant Problem Or Little Lambda Goes Back to the Future,''
Nucl.\ Phys.\ B {\bf 309} (1988) 493.


\bibitem{Beane:1997it}
S.~R.~Beane,
``On the Importance of Testing Gravity at Distances Less Than 1-Cm,''
Gen.\ Rel.\ Grav.\ {\bf 29} (1997) 945
[hep-ph/9702419].



\end{thebibliography}
\end{document}